\documentclass[12pt]{article}

\usepackage{setspace}
\usepackage{graphicx,subcaption,pstricks,pst-node,psfrag,amsthm,amssymb,amsmath,dsfont,bbm,setspace,algorithm,algpseudocode,tabularx,hhline,colortbl,float,tikz,multirow,color,caption}
\usepackage[round]{natbib}
\usepackage{hyperref}
\usepackage{indentfirst}
\usepackage{authblk}
\usepackage[all]{xy}

\usepackage{placeins}

\usepackage[english]{babel}
\setcitestyle{authoryear,open={(},close={)}}

\usepackage[normalem]{ulem}

\usepackage{tikz}

\definecolor{neur}{HTML}{8000FF}
\definecolor{sig}{HTML}{00FFFF}
\definecolor{nai}{HTML}{FF0000}

\begin{document}

\title{Brain Functional Connectivity Estimation Utilizing Diffusion Kernels on a Structural Connectivity Graph}

\author[a,b]{Nathan Tung}
\author[c,d,e]{Jerome Sanes}
\author[b]{Eli Upfal}
\author[d,e,f,*]{Ani Eloyan}
\affil[a]{\small Department of Applied Mathematics, Brown University, Providence, RI 02912}
\affil[b]{\small Department of Computer Science, Brown University, Providence, RI 02912}
\affil[c]{\small Department of Neuroscience, Brown University, Providence, RI 02912}
\affil[d]{\small Carny Institute for Brain Science, Brown University, Providence, RI 02912}
\affil[e]{\small Center for Neurorestoration and Neurotechnology, Veterans Affairs Providence Healthcare System, Providence, RI 02908}
\affil[f]{\small Department of Biostatistics, Brown University, Providence, RI 02903}
\affil[*]{\small Corresponding author: ani\_eloyan@brown.edu}

\date{}

\maketitle


\begin{abstract} 
Functional connectivity (FC) refers to the investigation of interactions between brain regions to understand integration of neural activity in several regions. FC is often estimated using functional magnetic resonance images (fMRI). There has been increasing interest in the potential of multi-modal imaging to obtain robust estimates of FC in high-dimensional settings. We develop novel algorithms adapting graphical methods incorporating diffusion tensor imaging (DTI) to estimate FC with computational efficiency and scalability. We propose leveraging a graphical random walk on DTI to define a new measure of structural connectivity highlighting spurious connected components. Our proposed approach is based on finding appropriate subnetwork topology using permutation testing before selection of subnetwork components comprising FC. Extensive simulations demonstrate that the performance of our methods is comparable to or better than currently used approaches in estimation accuracy, with the advantage of greater speed and simpler implementation. We analyze task-based fMRI data obtained from the Human Connectome Project database using our proposed methods and reveal novel insights into brain interactions during performance of a motor task. We expect that the transparency and flexibility of our approach will prove valuable as further understanding of the structure-function relationship informs future network estimation.
\end{abstract}


%\newpage
\doublespacing

\section{Introduction}\label{s:intro}

Brain functional connectivity (FC) is the field of study concerned with the estimation of functional interactions between brain regions. Functional magnetic resonance imaging (fMRI) has been widely used recently for quantifying FC (\cite{lindquist2008statistical}, \cite{penny2011statistical}, \cite{cribben2016functional}, \cite{ashby2019statistical}). In a typical FC study, one has a pre-defined set of seed regions often selected based on anatomical considerations or brain regions estimated by analytical methods such as independent component analysis \citep{calhoun2001method}. Each seed region typically comprises a collection of neighboring voxels or individual voxels. Next, one averages the blood-oxygen-level-dependent (BOLD) fMRI intensities in each region across space for each time point and obtains a time series $Y_{ir}(1), \ldots, Y_{ir}(T)$, for participant $i=1, \ldots, I$, in region $r=1,\ldots, R$, assuming the number of participants in the study is $I$, the number of regions of interest (ROIs) is $R$, and data are collected at $T$ time points. A simple approach for obtaining a brain FC map is computing the Pearson correlation between each pair of regions. As a result, one obtains an $R \times R$ matrix of correlations defined by $C_{i}$ for participant $i = 1,\ldots, I$. Other approaches  for computing brain FC include coherence analysis \citep{ombao2006coherence} and partial correlation estimates, among others. 

 Graph theory (\cite{chung2019brain}, \cite{bullmore2011brain}) is commonly used for modeling brain FC. In this context, one considers each region of interest as a vertex in an undirected graph defined as $G_i = (V_i, E_i)$, where $V_i$ represents the vertices of the graph and $E_i$ represents the edges constructed using the estimates of interactions between pairs of regions. We assume subject-specific data have been registered to a common template space and the same parcellation was used to extract the regions, resulting in a common set of vertices $V$. For instance, an edge may be drawn between two vertices if the correlation between time series corresponding to the vertex ROIs is higher than a pre-specified number. One may consider a weighted graph where the correlation between each pair of regions, defined as $C_{i}$ above, is the weight on the edge connecting vertices corresponding to the regions. Communicability (\cite{estrada2008communicability}, \cite{crofts2009weighted}) is another approach for estimation of graph network features as estimates of FC and is  computed as the shortest paths between nodes in $G_i$. The existing methods for estimation of FC have limitations in terms of computational scalability and reproducibility. Computationally efficient approaches for estimation of such graphs have been proposed, such as by estimating the precision matrix of fMRI time courses. However, many of these approaches may not be scalable to very large values of $R$. Another common issue of FC estimation approaches is lack of reproducibility of the resulting estimates, especially at the subject level. To address this problem, Gaussian graphical models have been implemented for estimating FC by incorporating structural connectivity (SC) \citep{higgins2018integrative} or population-level FC estimates \citep{varoquaux2010brain} via a prior distribution. We build on this work and propose biologically informed modeling of SC for estimation of FC in high dimensions. 

Accurate detection of subnetworks in the FC graph is imperative for correct network-level inference. FC can be estimated both at subject- and population-level. For instance, the average of subject-level maps $C_i$ is a possible estimator for population-level FC. Estimation of population-level FC has been considered for comparisons of brain functional organization between groups of interest and to use the resulting maps for better estimation of subject-level FC. For example, \cite{caffo2010two} proposed an approach for estimation of population-level FC by applying the singular value decomposition in two-stages to reduce the dimension of fMRI data and then investigated associations of the resulting FC estimates and cognitive impairment in Alzheimer's disease. Our proposed approach for FC estimation is presented as estimation procedure for population-level FC, however this approach can also be used to obtain subject-level estimates of FC that are more comparable between subjects and are less affected by subject-level noise. 

We propose novel biologically motivated and computationally efficient approaches to modeling SC as a prior for estimating the FC graph in high dimensions. Our proposed approach includes two major contributions: 1) innovative modeling of SC using heat diffusion processes, 2) novel scalable FC estimation framework and algorithms. First, we extend and implement a heat diffusion approach named HotNet, proposed by \cite{vandin2011algorithms} in the setting of genomic data analysis, to modelling SC. We provide a brief overview of diffusion tensor imaging (DTI)-based SC estimation and use DTI-based SC maps when estimating FC. HotNet is based on using ideas from heat diffusion (\cite{kondor2002diffusion}, \cite{lafferty2005diffusion}) to estimate the local topology of the SC graph. There is precedent for using diffusion processes in biology: \cite{qi2008finding} proposed a diffusion kernel-based method for analyzing genetic interactions in yeast. \cite{ma2007cgi} used heat diffusion to develop an approach for prioritizing genes using a combination of gene expression and protein-protein interaction networks. Our proposed modeling of SC is general and can be incorporated in other algorithms using SC as priors.

Next we propose a general framework for estimation of FC at both the subject- and the population-levels. We test two methods incorporating our novel notion of SC, one as an extension to a recent approach to FC estimation known as structurally informed Gaussian graphical model (siGGM, \cite{higgins2018integrative}) and the other - a simpler, more scalable model. We refer to our new treatment of siGGM as ``siGGM with Diffusion'' and our newly developed scalable model as ``Neuro-HotNet''. Using simulation studies, we show that both methods are more robust to noise than existing approaches. At the same time, when applied to analyze a database of fMRI scans from the Human Connectome Project (HCP), where study participants were performing motor tasks, both methods identify components expected to have involvement in motor movements, such as those containing the motor cortex and sensory-touch regions that surprisingly do not emerge with prior methods. In addition, we find that Neuro-HotNet is significantly more computationally efficient than known approaches, providing a practical tool for analyzing larger, more refined connectivity models.

\subsection{Outline}\label{ss:outline}

In Section \ref{s:methods}, we describe the novel use of the diffusion process applied to DTI data to estimate SC. We then show how the resulting SC estimate can be used to obtain FC estimates through either of two approaches: Neuro-Hotnet or siGGM with Diffusion. In Section \ref{s:sim}, we provide results of simulations to compare the estimation accuracy and runtimes of the different approaches discussed in this paper. 
% We also explore edge-cases/scenarios that give further insight into the performance of Neuro-Hotnet. 
Section \ref{s:application} details the results of applying three approaches (Neuro-Hotnet, siGGM with Diffusion, and SC Naive) to estimate FC during the execution of a motor task. These results are compared and discussed in the context of neurological implications of the task performed by the participants. Section \ref{s:discussion} includes summary of the proposed methods and final conclusions.

\section{Methods}\label{s:methods}

In this section, we provide background and visualizations to describe the diffusion process used to estimate SC. We then present our proposed Neuro-HotNet approach for selecting connected subnetworks with a hypothesis testing procedure. Finally, we give an overview of the siGGM approach and show how it can be modified to use our diffusion-based SC.

\subsection{Structural Connectivity}\label{ss:DTI}

\begin{figure}[ht]
\begin{center}
\includegraphics[scale=0.55]{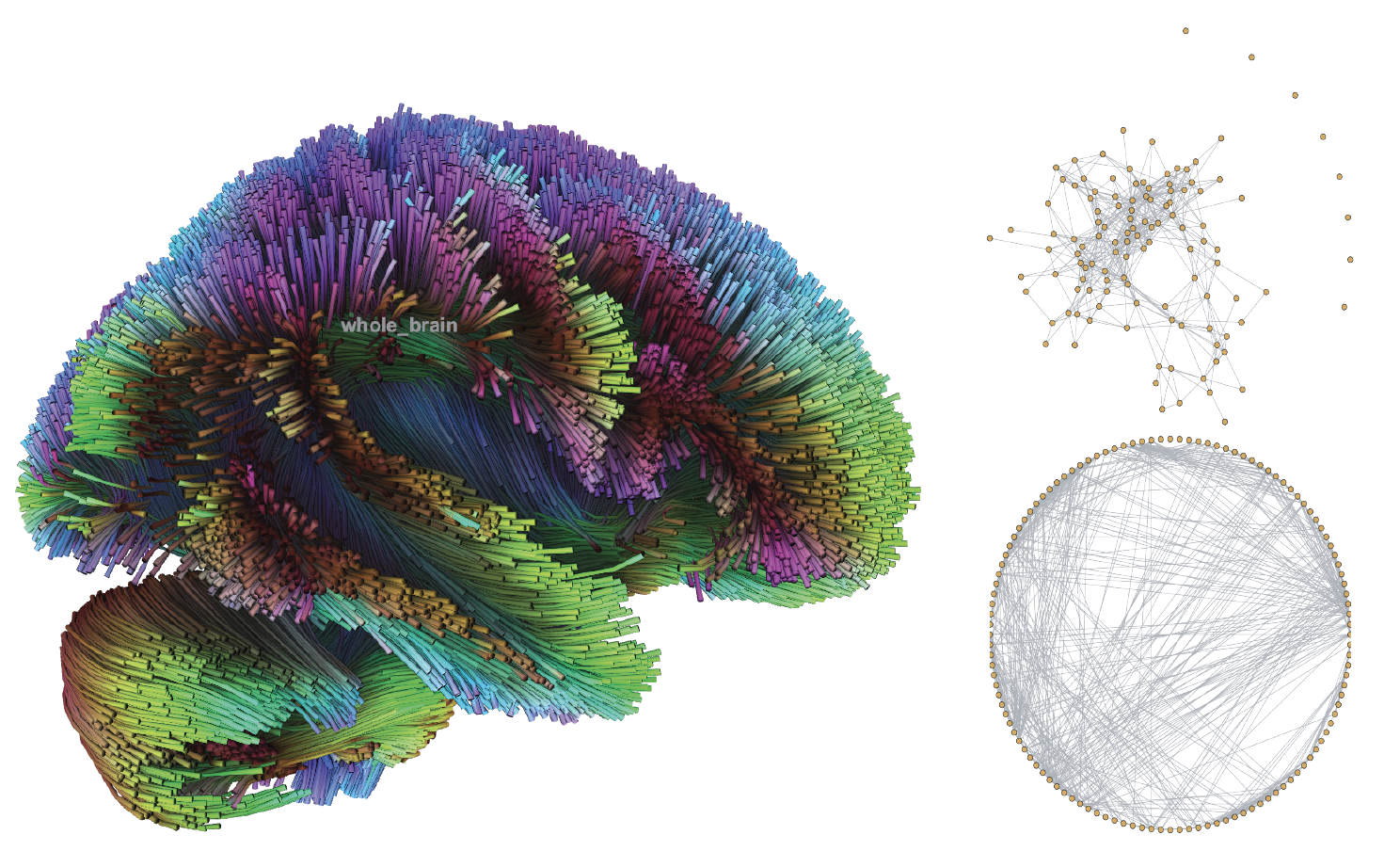} 
\caption{Left: white matter fiber tracts estimated using the HCP template DTI plotted in MNI space. Right: SC graph estimated using  the AAL parcellation of the DTI data shown on the left. The graph is presented using a random node structure (top) and a nodes on a circle structure (bottom).}\label{fig:tracts}
\end{center}
\end{figure}

DTI is a widely used MRI technique where the structure of biological tissues in the brain is characterized using anisotropy, magnitude, and anisotropic orientation of water diffusion \citep{basser1994mr}. DTI has been used in recent years to obtain estimates of SC by tracing the white matter fiber tracts \citep{mori2002fiber}. An example of white matter fiber tracts computed using the HCP template DTI dataset is shown in Figure \ref{fig:tracts}. For a given parcellation of the brain, SC between any two pairs of regions  can be computed using the number of fiber tracts connecting the regions normalized by tract lengths, parcel size, e.g. surface area or volume \citep{chung2019brain}. Assuming the parcellation includes $R$ regions, we obtain an $R \times R$ structral connectivity matrix for a participant. Figure \ref{fig:tracts} shows the SC graphs computed by using the AAL parcellation of the HCP template DTI, where only the strongest connections (highest number of tracts) are shown. When we consider estimation of a subject-specific FC map, we assume the same parcellation is used for DTI and fMRI data, similarly, when considering population-level FC estimation, we assume the same parcellation is used for DTI and each subject-specific fMRI data. We use the notation $V$ to denote the vertices corresponding to these common regions and $R$ is the total number of regions.

\subsubsection{Heat Diffusion}\label{s:diff}

We implement a ``heat'' diffusion process using the SC map from DTI based on the assumption that the strength of interactions between two nodes depends on the neighborhood topology of these nodes. Let $G_S = (V, E_S)$ denote the graph corresponding to the structural brain map with edge and node sets $E_S$ and $V$ respectively, $|V| = R$. The process of obtaining the SC map can be executed using unweighted (binarized) edges, showing whether there is a large enough number of tracts between two nodes to conclude they are structurally connected, or edges weighted by the number of tracts between all pairs of nodes \citep{chung2019brain}. Given the resulting binary SC map, our implementation of the heat diffusion estimation for the unweighted graph mirrors that proposed by \cite{vandin2011algorithms}. However, we extend this approach by proposing a weighted method to take advantage of the additional information on number of tracts between any two ROIs available from DTI. First, we normalize \citep{qi2008finding} the DTI SC estimate to obtain $M' = D^{-1/2}MD^{-1/2}$, where $M$ is the adjacency matrix representation of the graph $G_S$ and $D$ is the diagonal matrix of row sums of $M$. For the unweighted graph these are the binary adjacency matrix and diagonal degree matrix, while for the weighted graph these can simply be thought of as weighted analogs. The elements of the adjacency matrix in both cases are transformed as follows. 
\begin{equation}\label{eq:normprocess}
    M'_{r_1r_2} = \frac{M_{r_1 r_2}}{\sqrt{\left(\sum_{m} M_{r_1m }\right)\left(\sum_{m} M_{m r_2}\right)}}
\end{equation} 

This process is analogous to scaling the values by node degrees (either the typical or weighted) with the result that every path is weighted according to degrees of all the nodes on the path. Let $L$ denote the Laplacian matrix of $G_S$ defined as $L = -M' + D'$, where $D'$ is the diagonal matrix of row sums of $M'$. We first compute the influence of each node $r$, where $r \in \{1,\ldots,R\}$, on the rest of the nodes in the graph. We consider the process where a certain number of random walkers are placed at node $r$ and they diffuse across the graph at a given rate $\gamma$. Let $f^r(t) = (f^r_1(t), \ldots, f^r_n(t))$ denote the distribution of heat among the $R$ nodes. Then, the equation describing the dynamics of this process is given as follows (\cite{qi2008finding}, \cite{vandin2011algorithms}).
\begin{equation}\label{eq:differencial}
\frac{\partial f^r(t)}{\partial t} = - (L + \gamma I) f^r(t) + b^r u(t)
\end{equation}
where $b^r$ is a vector of zeros except for 1 at the $r^{th}$ location and $u(t)$ is the unit step function. As $t \rightarrow \infty$, the equilibrium of (\ref{eq:differencial}) is given by $f^r = (L+ \gamma I)^{-1} b^r$.

The influence of all node pairs may be computed simultaneously as $f = [(L+ \gamma I)^{-1} ]^T$,  where $f_{r_1r_2}$ is the influence of node $r_1$ on node $r_2$. After the influence matrix is computed it is once again normalized, as in \cite{higgins2018integrative}, so that each entry $r_1,r_2$ can be thought of as the probability that a tract with an endpoint in region $r_2$ has its other endpoint in region $r_2$. Formally, we obtain the influence matrix $G(\gamma)$ such that
\begin{equation}\label{eq:normalizing}
    G(\gamma)_{r_1r_2} = \frac{1}{2}\left[\frac{f_{r_1r_2}}{f_{r_1*}}+\frac{f_{r_2r_1}}{f_{r_2*}}\right],
\end{equation}
where $f_{r_1*}$ and $f_{r_2*}$ are the sums of rows $r_1$ and $r_2$, respectively. In practice, this step functions similarly to the degree-normalizing step before inverting the Laplacian, i.e., reducing influence on high degree nodes.

\begin{figure}[ht]
\centering
    \includegraphics[scale=0.55]{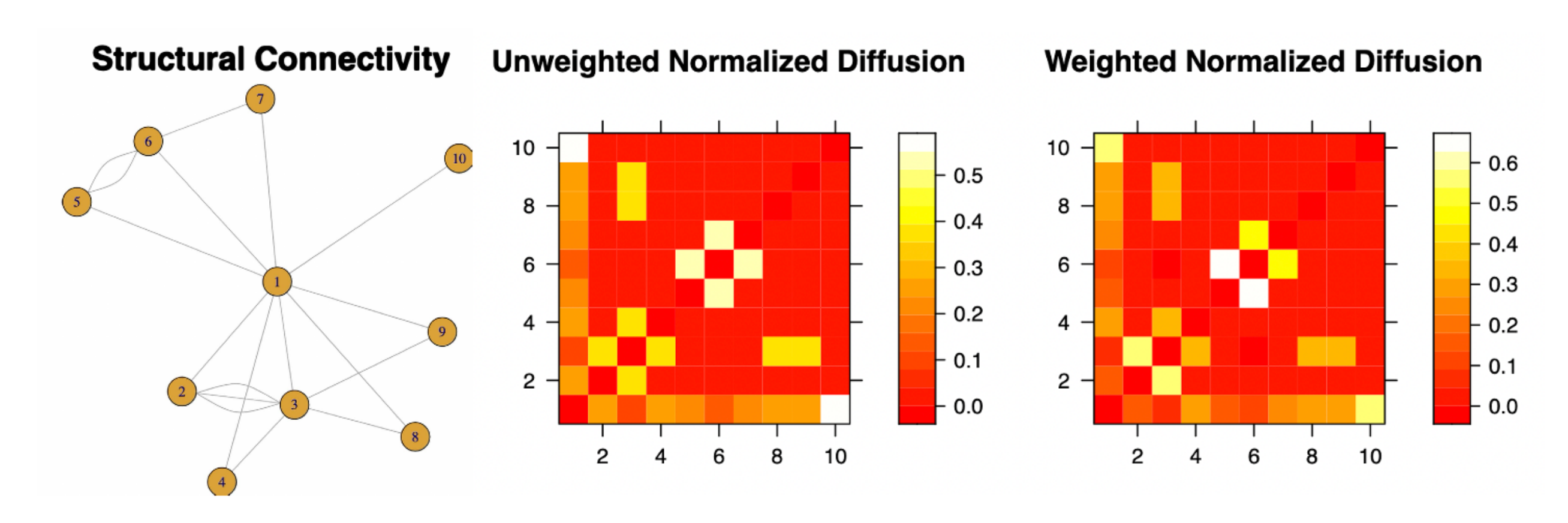}
\caption{Left: An example undirected graph with 10 nodes. Multiple edges indicate relative weighting. Results of heat kernel diffusion process using the binary SC in the middle and the weighted SC map on the right. Here normalization refers to (\ref{eq:normalizing}), while the process described in (\ref{eq:normprocess}) is executed for all depicted heatmaps.}
\label{fig:diff}
\end{figure}

In Figure \ref{fig:diff}, the diffusion process is illustrated using a toy example. Influence from node 1 is spread among its many neighbors, whereas the influences between 5, 6, and 7, for example, are strong, since they have relatively few other neighbors. It is clear that weighting the diffusion process appropriately incorporates the relative weights of edges (2-3, 5-6). The effect of node 1's influence due to its many neighbors is visible in the maps Figure \ref{fig:diff} middle and right, as illustrated by the strength of connection on the edge 1-10, given that node has no other neighbors, as compared to the lower strength of connection between 1 and the rest of its edges. A pseudocode outline of the diffusion scheme implemented for DTI data is illustrated in Algorithm \ref{alg:diff}.

\begin{algorithm}
    \caption{Diffusion Algorithm}
    \textbf{Input:} SC matrix $M$ and flow-rate parameter $\gamma$ \\
    \textbf{Output:} Influence Graph $G(\gamma)$
    \begin{algorithmic}[1]
        \State Construct diagonal node-degree matrix $D$, with $D_{r_1r_2} = \mathbbm{1}_{r_1 = r_2} \times \sum_{x=1}^R M_{r_1x}$;
        \State $M' \gets D^{-1/2}MD^{-1/2}$;
        \State Construct $D'$, with $D'_{r_1r_2} = \mathbbm{1}_{r_1 = r_2} \times \sum_{x=1}^R M'_{r_1x}$;
        \State $L \gets -M' + D' + \gamma I$;
        \State $f \gets [L^{-1}]^T$;
        \For{$r \in \{1,\ldots,R\}$}
            \For{$j \in \{1,\ldots,R\}$}
                \State $G(\gamma)_{r_1r_2} = average\left(\frac{f_{r_1r_2}}{\sum_{x=1}^R f_{r_1x}}, \frac{f_{r_2r_1}}{\sum_{x=1}^R f_{r_2x}}\right)$
            \EndFor
        \EndFor
        \State \Return $G(\gamma)$
    \end{algorithmic}
    \label{alg:diff}
\end{algorithm}

\begin{figure}[ht]
\centering
    \includegraphics[scale=0.4]{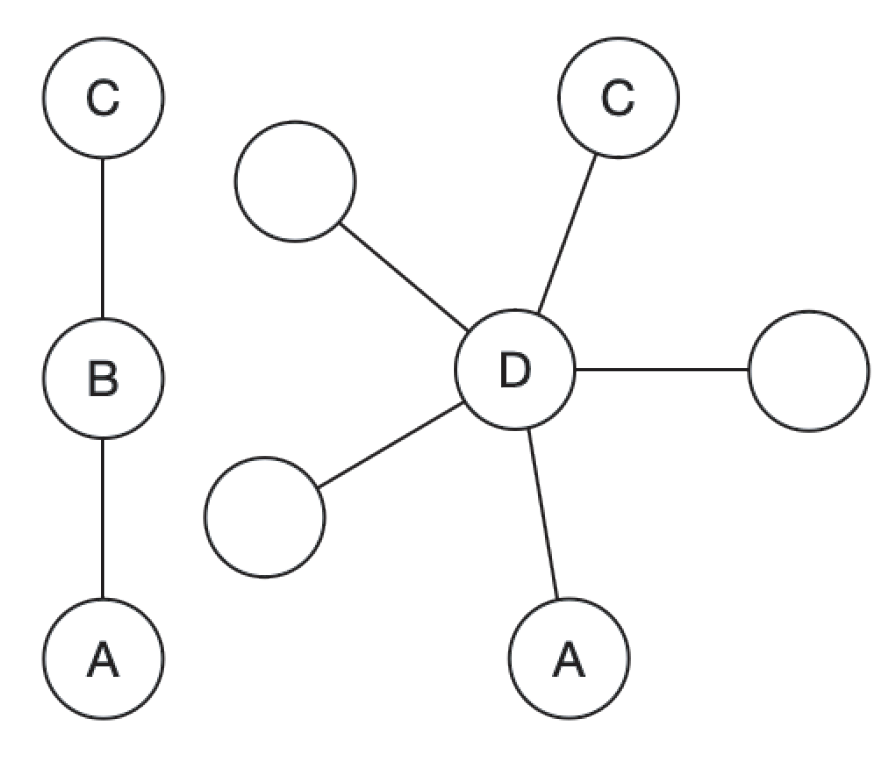}
    \caption{Chain versus star graph illustrating the effect that the number of connections of a node has on its influence.}
    \label{fig:chainstar}
\end{figure}

The idea behind this approach parallels the following description on social network modeling. We consider modeling associations between individuals a social networks presented in Figure \ref{fig:chainstar}. Suppose individual D in a social network is connected to a large number of individuals, while individual B is connected to few others in the network. If A is connected to C through D this connection may not be of interest, as D is connected to many individuals implying that many of their connections are superficial: the influence between A and C would become down-weighted. If A is connected to C through B their connection potentially has more interest as B's few neighbors imply its connections are substantial: the influence between A and C would then become up-weighted. Replacing people in this canonical example with ROIs gives the reasoning behind our proposed diffusion modeling. In this approach, not only the number of connected components of two ROI's, but also those of all nodes in the path connecting them are taken into account. 

\subsection{Neuro-Hotnet}\label{s:Neuro-Hotnet}

In this subsection, we discuss selection of subnetworks of interest using the SC framework for hypothesis testing for identifying the subnetworks including significant functional connections estimated by fMRI. Finally, we show how our proposed diffusion SC approach can be incorporated in methods where prior information can be used for estimation of SC by using the siGGM as an example. 

\subsubsection{Selecting Networks}\label{sec:usingsc}

In this subsection, we discuss the selection of candidate networks. The proposed framework is based on the key assumption that a network is only of interest if it is both strongly connected in the structural data as well as significantly connected in the fMRI data. This approach has precedent in the literature \citep{higgins2018integrative}. To avoid invalid Type I error rates due to selective inference (\cite{fithian2014optimal}, \cite{gao2022selective}), we use SC to select possible networks of interest and use only fMRI data to test for significance of these networks. Hence, the model selection step of the procedure is performed independently of the testing procedure in the second step of the algorithm. 

In order to identify a set of strongly connected subnetworks in the SC graph we induce sparsity by removing all edges with weight below a tuning parameter $\delta$, which will be referred to as the threshold. The choice of $\delta$ will be discussed further in Section \ref{sec:comptest}. Let $H(\delta)$ denote the resulting thresholded graph. The set of candidate subnetworks includes all subnetworks of $H(\delta)$ of at least 3 nodes. We do not consider pairs of nodes as these just constitute a single edge, and identifying single highly weighted edges not only defeats the purpose of the diffusion process but is also not very meaningful in the context of FC estimation.

\subsubsection{Testing of Individual Components}\label{sec:comptest}

We implement the combined hypothesis testing procedure (Chapter 9 of \cite{efron2012large}, \cite{wang2019combined}) proposed in the context of gene enrichment analysis to estimate significant subnetworks of FC. In gene enrichment analysis, it is often known which gene groups form pathways relating to various functions. Hence, testing for non-null enrichment in each of these pathways, instead of testing for mutations in individual genes is preferred for discovering existing effects that may be missed by individual hypothesis tests even with application of corrections for multiple comparisons. Extending this idea to FC estimation, we propose using the subnetworks obtained from the diffusion enhanced graph to identify the subset of subnetworks that are functionally connected at the population-level. Let $H_{S}(\delta) = \{H_1, \ldots, H_L\}$ define the set of all connected components of $H(\delta)$, where $L \leq 2^R$.

We perform a combined hypothesis test for each subnetwork in $H_S(\delta)$ to test whether FC of the subnetwork is significantly different from zero. For a given subnetwork $H_l \in H_S(\delta)$, let $V_H$ define the set of nodes in $H_l$. By our definition, $C_{i,r_1r_2}$ denotes the functional connectivity between regions $r_1,r_2 \in V_H$ for participant $i$ measured using the correlation of average BOLD signals observed in the corresponding regions. Let $C$ define the $R \times R$ population-level true functional connectivity that we are interested in estimating. Then, following \cite{efron2012large}, we implement the model $C_{i,r_1r_2} = C_{r_1r_2} + \epsilon_{i,r_1r_2}$, where $\epsilon_{i,r_1 r_2} \sim N(0, \sigma^2)$ are random noise independent across study participants for each pair of ROIs. Potential associations between error components are ignored in this model, however, future research may investigate effects of groupings in study participants resulting in structured error variance components. We are interested in testing the following combined hypothesis
 \begin{equation}\label{eq:hyptest}
     H_0: C_{r_1r_2} = 0, \forall r_1, r_2 \in V_H \mbox{ vs } H_a: C_{r_1r_2} \neq 0, \mbox{ for some } r_1,r_2 \in V_H
\end{equation}
In order to test this hypothesis, we use a permutation testing approach. Under the null hypothesis, the correlations $C_{i}$ are randomly permuted thereby eliminating any association of FC and true regional assignment. Specifically, for each study participant $i$ we permute the rows of $C_i$ uniformly at random to generate a corresponding sample under the null hypothesis. Let $Z_{ir_1r_2}$ define the z-scores of the sample correlations $C_{ir_2r_2}$ computed using the Fisher z-transformation. With the following test statistic \citep{efron2012large}
\begin{equation}\label{eq:indivtest}
    S(Z_H) = \frac{1}{s} \sum_{i=1}^I \sum_{r_1,r_2 \in V_H} Z_{ir_1r_2} 
\end{equation}
we compute the p-value to perform \ref{eq:hyptest} as follows. 
\begin{equation}\nonumber
    p = \frac{\# \{S(Z_H) \le S(Z_{0_H}) \}}{B}
\end{equation}
where $Z_0$ are the z-scores of sample correlations in the samples generated under the null hypothesis assumption, while $B$ is the number of permutation samples. The result is a population-level p-value for each connected component in $H_S(\delta)$.

It remains to discuss the selection of the tuning parameter $\delta$. The role of $\delta$ is to determine the relative sizes of discovered components. Increasing $\delta$ can only remove edges. Hence, for a given subnetwork, increasing $\delta$ either shrinks the component or splits it into smaller subnetworks without changing the location of the subnetwork itself. The higher values of $\delta$ result in smaller, more specific subnetworks. Hence, if the goal of the investigator is to select larger subnetworks, then smaller values of $\delta$ would be recommended. Future research may consider potentially estimating the value for $\delta$ using the DTI data, however, at this stage the selection of $\delta$ is considered as a tuning parameter selection problem based on the goals of the user of the proposed algorithms and biological considerations. The pseudocode for all of Neuro-Hotnet is given in Algorithm \ref{alg2}.

\begin{algorithm}
    \caption{Neuro-Hotnet Algorithm}
    \textbf{Input:} Average regional cross-correlations $\tilde{P}_k$ for each of N participants $\tilde{P}$, influence graph from diffusion $G(\gamma)$, significance level $\alpha$, and threshold $\delta$\\
    \textbf{Output:} Connected components and their p-values
    \begin{algorithmic}[1]
        \State $H(\delta) \gets G(\gamma)$ with values $< \delta$ set to 0; \label{start}
        \State $C \gets$ connected components of $H(\delta)$ \label{end}
        \State Initialize vector $p$ of p-values to 0
        \ForAll{$c \in C$}
            \State Initialize vector $\mu_0$ of mean edge weights under null
            \For{$k \in \{1,\ldots,N\}$}
                \State Shuffle $\tilde{P}_k$ (null distribution), call this $\tilde{P}_{k_0}$
                \State $\mu_0[k] \gets \frac{1}{|c|}\sum_{i,j \in c} |\tilde{P}_{k_0}[i,j]|$
            \EndFor
            \State $p_c \gets$ p-value from two-sample paired t-test. Sample 1: $\frac{1}{|c|}\sum_{i,j \in c} |\tilde{P}_k[i,j]|, \forall k \in \{1,\ldots,N\}$. Sample 2: $\mu_0$
        \EndFor
        \State $C^* \gets \{c \in C: p_c < \frac{\alpha}{|C|}\}$ 
        \State\Return $(C^*,p|_{C^*})$
    \end{algorithmic}\label{alg2}
\end{algorithm}

\subsection{siGGM with Diffusion}

In this subsection, we discuss incorporating our proposed modeling of the diffusion processes in an example FC estimation approach using Gaussian Graphical modeling. Methods such as siGGM, as proposed by \cite{higgins2018integrative}, which use Bayesian Gaussian graphical models in lieu of multiple hypothesis testing have shown promise in structurally-informed FC estimation. Given $Y_i(t), t=1,\ldots T$, representing the vector of BOLD signals at all ROIs at time $t$, siGGM assumes the statistical probabilistic model $Y_i(t) \sim N(0, K_i^{-1}),$ where $K$ is the inverse covariance matrix. Graphical lasso \citep{friedman2008sparse} is a commonly used regularization method for estimation of precision matrices by incorporating sparsity as follows.
\begin{equation}\label{eq:higobjective}
\hat{K_i} = argmax_{K_i \in M^{+}_R} \log \det (K_i) - trace( \Sigma_i K_i) - \lambda \sum_{j \leq k} | K_{ijk} |
\end{equation}
where $M_{R}^{+}$ is the cone of $R \times R$ symmetric positive definite matrices, while $\Sigma_i$ defines the sample covariance matrix. In the past decade, population level FC estimates have been used to obtain more precise estimates of subject-level covariance.  Bayesian graphical models \citep{wang2012bayesian} are a convenient modeling approach implemented for incorporating population level maps using prior distributions. \cite{higgins2018integrative} proposed using anatomical knowledge (SC) as a prior for estimation of subject-specific FC using the following prior density for the inverse covariance.
\begin{equation}\nonumber
P(K_i | \lambda) = Z^{-1}_{\lambda} \prod_{r=1}^R  Exp(K_{irr} | \lambda) \prod_{j < r} DE(K_{i j r} | \lambda) I(\Sigma_i \in M_{R}^{+})
\end{equation}
While the prior densities implemented for incorporation of SC are given as follows.
\begin{align*}
P(K_i | \lambda) & = Z^{-1}_{\lambda, \nu} \prod_{r=1}^R  Exp(K_{irr} | \frac{\nu}{2}) \prod_{j < r} DE(K_{i j r} | \nu \lambda_{jr}) I(\Sigma_i \in M_{R}^{+}) \\
P(\lambda | \mu, \eta) & = Z_{\lambda, \nu}  \prod_{j < r} LN ( \mu_{jr} - \eta e_{jr}, \sigma_{\lambda}^2)
\end{align*}
The algorithm then finds the maximum \textit{a posteriori} (MAP) estimate by maximizing the posterior log-likelihood:
\begin{align*}
&\dot{l}(\boldsymbol{\Theta})=-\frac{\mathrm{T}}{2} \log |\boldsymbol{\Omega}|+\frac{1}{2} \operatorname{tr}(\mathrm{S}|\boldsymbol{\Omega}|)+\nu \sum_{\mathrm{j}<\mathrm{k}} \mathrm{e}^{\alpha_{\mathrm{jk}}}\left|\omega_{\mathrm{jk}}\right|+\sum_{\mathrm{j}<\mathrm{k}} \frac{\left(\alpha_{\mathrm{jk}}-\left(\mu_{\mathrm{jk}}-\eta \mathrm{p}_{\mathrm{jk}}\right)\right)^{2}}{2 \sigma_{\lambda}^{2}} \\
&-\left(\mathrm{a}_{\eta}-1\right) \log (\eta)+\mathrm{b}_{\eta} \eta+\sum_{\mathrm{j}<\mathrm{k}} \frac{\left(\mu_{\mathrm{jk}}-\mu_{0}\right)^{2}}{2 \sigma_{\mu}^{2}}-\operatorname{plog}\left(\frac{1}{2} \nu\right)+\frac{1}{2} \nu \sum_{\mathrm{k}=1}^{\mathrm{p}} \omega_{\mathrm{kk}}
\end{align*}

While the details of interpretation and choices of prior parameters can be found in \cite{higgins2018integrative}, note the role of $\nu$ in the log-likelihood to penalize edge weights ($\omega$) and induce sparsity; $\nu$ roughly plays the role of $\delta$ in Neuro-Hotnet. The choice of $\nu$ exhibits exactly the same tradeoffs as the choice of $\delta$. Estimates perform similarly not changing locations in the brain based on $\nu$, but rather shrinking and growing in size. $\nu$ can be used similarly to $\delta$ to find subnetworks of a desired size.

We propose incorporating the diffusion process outlined in Section \ref{s:diff}, to obtain subject-specific FC estimates while using our proposed heat diffusion SC within the framework of siGGM. This flexible framework allows for estimation of FC by replacing the SC by the influence graph from the diffusion process. We assume BOLD fMRI signals are normalized, i.e. by subtracting the mean of the time course and dividing by the standard deviation, to ensure that the observations and SC are on the same scale. 

Note that the above approach is aimed at estimation of FC at the subject-level. To obtain a population-level estimate of FC given a sample of fMRI scans, we propose the following approach. For each participant's fMRI data, we calculate the sample covariance matrix across regions over all time points. These covariance matrices are then averaged and used as input in siGGM to obtain a population-level FC estimate. Let $C(\nu)$ be the maximum posterior log-likelihood siGGM estimate of the population-level $R\times R$ covariance matrix incorporating all normalized BOLD signals $G(\gamma)$. $\nu$ refers to the tuning parameter controlling the network's overall sparsity and must be tuned to balance number of discoveries with sensitivity. The set of connected components of $C(\nu)$, denoted $S(\nu)$ in the sequel, are returned as strongly connected subnetworks. When implementing the siGGM method, we used the source code provided by the authors (see \cite{higgins2018integrative}) and the default configurations of the  parameters to adjust the relative impact of structural influence and observations.

\begin{figure}[ht]
    \centering
    \includegraphics[scale=0.55]{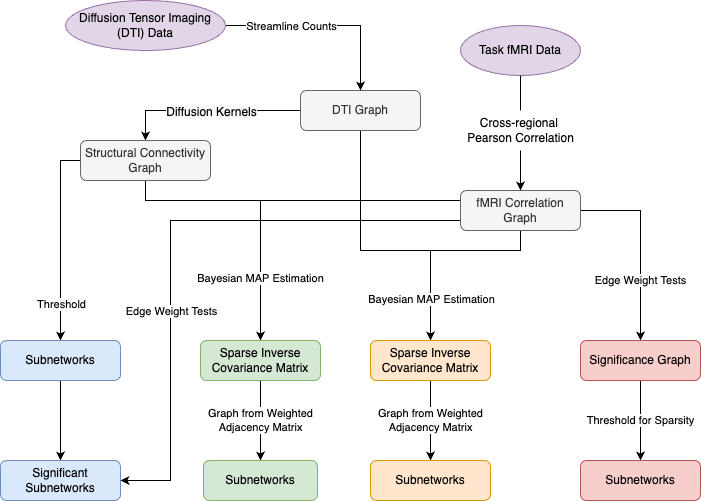}
    \caption{Overview of four approaches to FC estimation and the use (or lack thereof) of SC in each. Novel methods are depicted on the left with Neuro-Hotnet in blue and siGGM with Diffusion in green, while the currently used siGGM and SC naive are shown in orange and red respectively.}
    \label{fig:pipeline}
\end{figure}

Figure \ref{fig:pipeline} presents all methods for FC estimation used in this paper, including the two novel approaches discussed in this section. The SC naive method is a commonly used approach implemented to evaluate the comparative performance of our proposed methods for FC estimation. As the name suggests, this method does not incorporate any structural or DTI data and solely identifies subnetworks through correlations derived from fMRI. First, a graph is created by testing whether the correlations between the time courses at each pair of nodes (regions) across all participants are different from the  average correlation between all pairs of regions. In other words, the nodes of the resulting graph consist of all brain regions of interest, while two nodes are connected with an edge if the correlations between time courses of these nodes across the study participants are significantly different from the average correlation between all pairs of regions across all participants. The hyperparameter $\epsilon$ is used as the threshold for the p-values. The connected components of the resulting graph are returned as strongly connected subnetworks.

\section{Simulations: Comparing and Examining Methods}\label{s:sim}

In this section, we first perform a realistic simulation using the parameters estimated from the HCP data to compare the performance of Neuro-Hotnet, siGGM with and without Diffusion, and the SC naive approach to estimate a known FC graph with fMRI data generated with additive noise from a structural prior. Then we compare runtimes of both novel algorithms (Neuro-Hotnet and siGGM with Diffusion) and how the computation times scale with graph size.

\noindent {\bf Study 1: Data parameters with noise.} To compare the performance of the SC naive approach, raw siGGM (no diffusion), siGGM with Diffusion, and Neuro-Hotnet, we generate sample SC graphs with $R=120,300,500$ nodes and compare the accuracy of each method in recovering these graphs. For each $r \in \{120,300,500\}$, an SC graph $M_r$ is generated with density 0.3 ($30\%$ of all pairwise edges are nonzero) and edge weights generated from the $uniform(0,1)$ distribution. Next all components of size $< 8$ are removed for ease of visualization/comparison and to mimic the results from the HCP analyses in the 120 ROI case. For each of these $M_r$ we then generate $n=308$ samples from a Gaussian distribution under the assumption that the row means, standard deviations, and dimensions mirror those of the HCP data, but correlations are equal to $M_r$ scaled to the nearest correlation matrix $c_{M_r}$. With $M_r$ denoting the $r \times r$ symmetric matrix representation of the graph edges,
\begin{align}
\begin{gathered}
    \rho_{M_r} = \frac{M_r}{\max |M_r|}, \\
    \rho_{M_{r_{ij}}} = 1, \forall i,j \in \{1,\ldots,r\} \land i=j, \\
    \Sigma_{M_r} = \mbox{nearPD}(S \times S^T \odot c_{M_r}),
\end{gathered}
\end{align}
where $\Sigma_{M_r}$ is the nearest covariance matrix used to generate the sample. The nearest positive definite matrix for a given matrix $A$ is computed using the method by \cite{higham2002accuracy} that finds the positive semidefinite matrix with unit diagonal minimizing $\{\lVert A-X \rVert: X \mbox{ is a correlation matrix}\}$, where $\lVert \cdot \rVert$ is a weighted Froebenius norm. Noise was then similarly generated for each participant from a multivariate Gaussian distribution with mean 0 and standard deviation 120 (for reference the mean signal strength in the fMRI data is 9,600) for all regions, all independent of each other. This simulation mirrors the structure of the HCP data, because each $M_r$ represents the anatomical brain structure that gives rise to the observations that are then contaminated by noise due to participant differences and measurement error.

\begin{figure}[ht]
    \centering
    \includegraphics[scale=0.4]{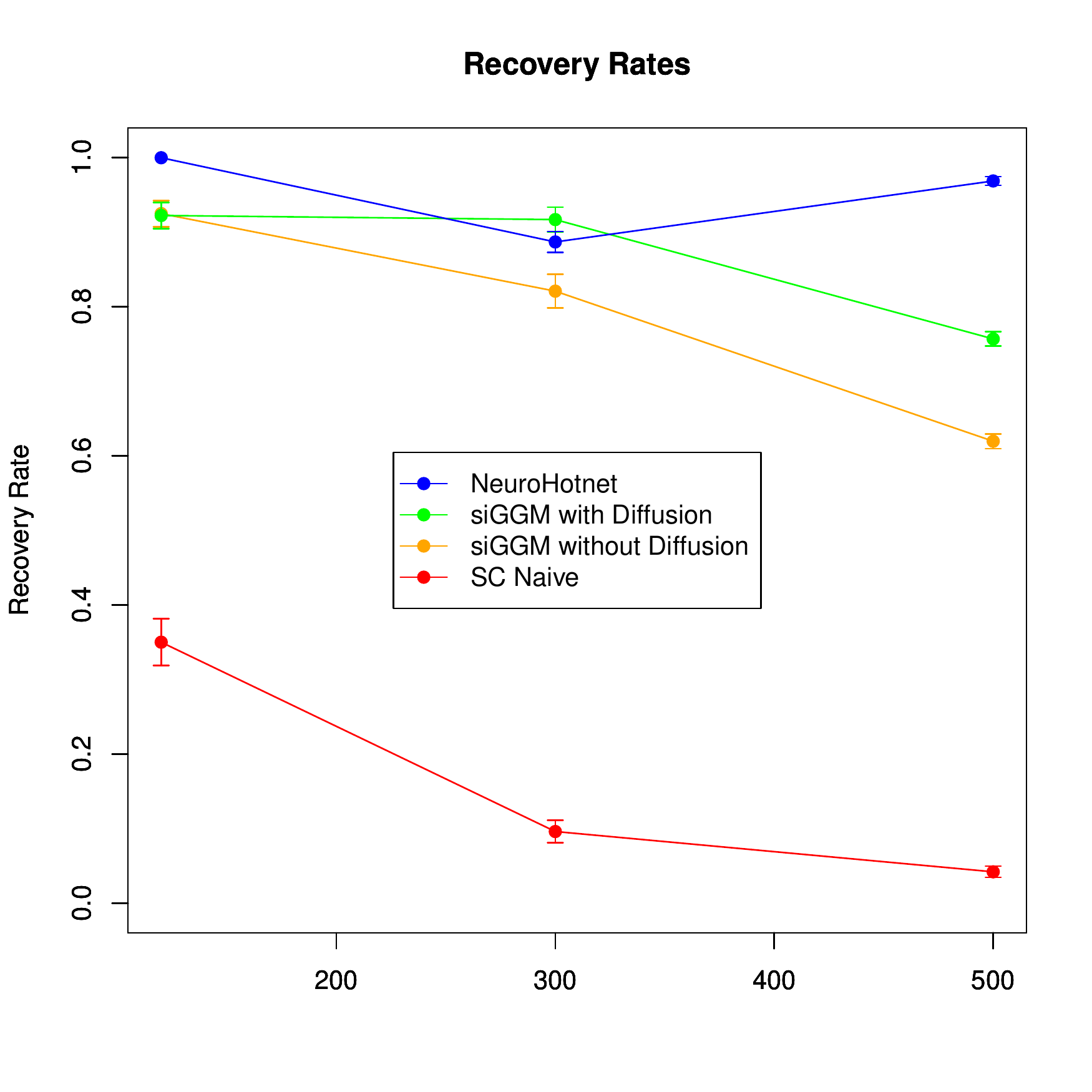}
    \includegraphics[scale=0.34]{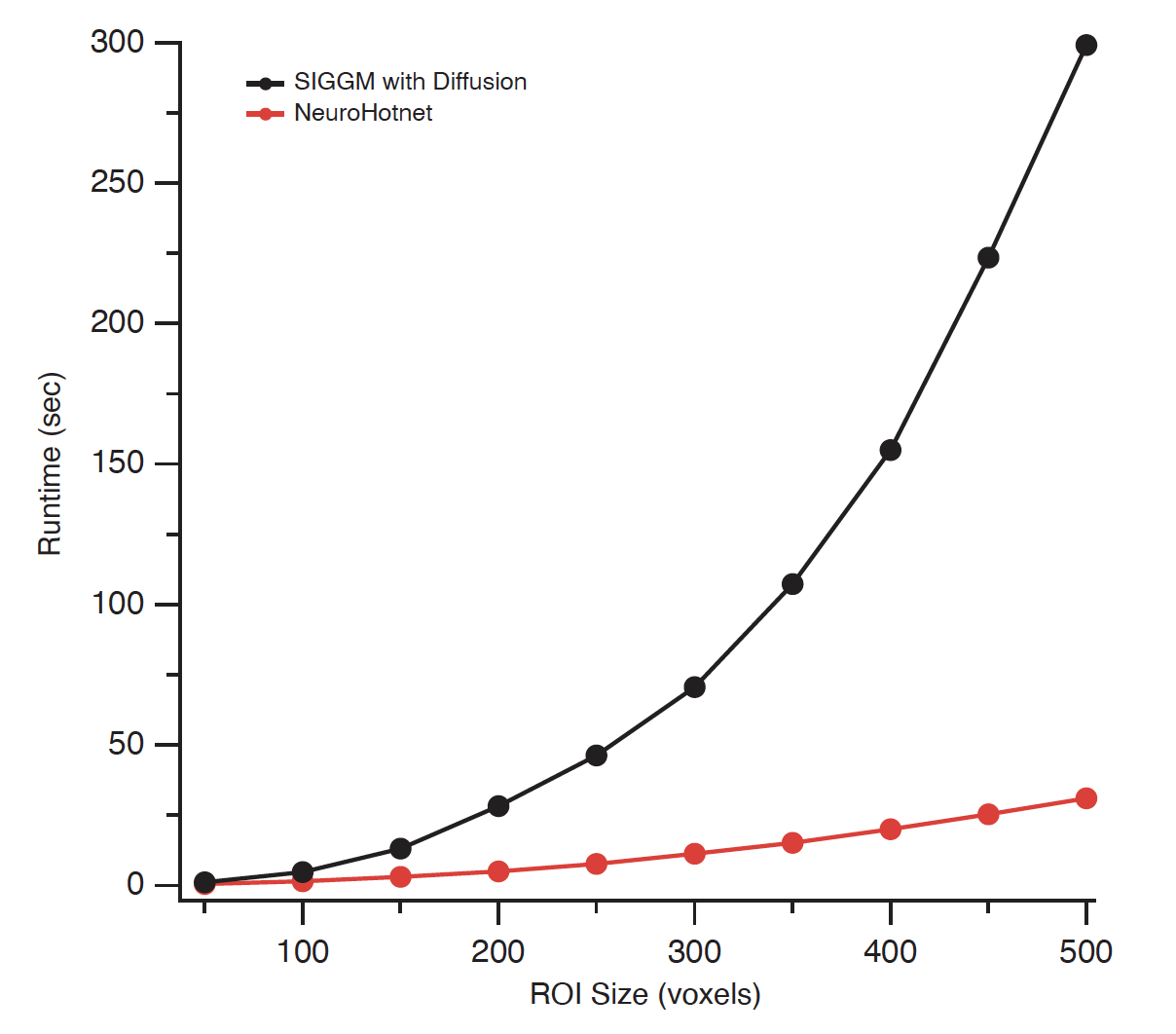}
    \caption{Left: Comparison of each method's average recovery rate for components of randomly generated SC graphs of varying sizes. 200 high-noise trials were run for each graph. Vertical bars represent $95\%$ confidence intervals. Rates for each graph are calculated as the proportion of components in the graph that have a match in a method's estimates, where two components are considered a match if they differ on less than 3 regions (any 2 or less errors, false positives or missing regions, are tolerated). Right: Runtime for both novel algorithms plotted against number of ROIs (i.e. number of nodes in the graph).}
    \label{fig:recrates}
\end{figure}

Figure \ref{fig:recrates} shows differences in estimation accuracy over 200 trials for each method for each graph. All SC informed methods perform significantly better than SC naive, reaching over $60\%$ recovery at all sizes at this level of noise despite the low error tolerance (see Figure \ref{fig:recrates} caption) used to define a successfully estimated component. NeuroHotnet performs particularly well for the largest graph size, beating out both siGGM methods with significantly shorter runtime. It is also the only method for which we know p-values can be accurately computed, as any attempt to carry out a similar hypothesis test for the results of siGGM runs into the issue of selective inference: components were selected using both data sources, so testing component significance with the same data would lead to inflated p-values. In this sense we are guaranteed significance of Neuro-Hotnet's results but not siGGM's. For both methods utilizing the diffusion process $\gamma$ was set at 30 (same value used for modeling the HCP data). Both methods using the diffusion process perform uniformly at least as well and for larger graphs quite a bit better than siGGM without diffusion. In general, there is a drop in recovery rate with increasing number of ROIs in all algorithms except for Neuro-Hotnet.

All algorithms implemented in these analysis have tuning parameters, that were chosen to obtain best performance in each ROI setting used in the simulation studies. An advantage for Neuro-Hotnet is that the threshold value $\delta$ was more robust to graph size (number of ROIs) than siGGM's tuning parameter $\nu$. A single value of $1.8\times10^{-3}$ for $\delta$ gave the best recovery in all cases. For 120 ROI the sparsity parameter $\nu$ was set to $2.5\times10^{-4}$ for siGGM with and without diffusion. For 300 and 500 ROI both were set to $1.5\times10^{-4}$. Finally, $8\times10^{-4}$ was used as the threshold parameter $\epsilon$ for all ROI's when implementing SC Naive.

\noindent {\bf Study 2: Runtimes.} We compare computation times of Neuro-Hotnet against those of siGGM with Diffusion across 10 different graph sizes (number of ROIs) to determine how their runtimes may scale with advancements in data collection methods resulting in finer parcellations of the brain and thus higher dimensional graphs. For each graph size, a simulated SC graph is generated as in Study 1 with density 0.3 and uniformly distributed edge weights. Each graph is then used as a prior to generate data for $n=308$ participants from a Gaussian distribution as in Study 1. Both algorithms are then applied 10 times for each graph size on an AMD Ryzen 5 3600 6-Core processor to attain the average runtimes plotted in Figure \ref{fig:recrates}. This illustrates Neuro-Hotnet's superior scalability due to its much simpler model for estimation.

\FloatBarrier
\section{Application: Estimation of FC During a Motor Task}\label{s:application}

We modeled the task-fMRI data publicly available by the Human Connectome Project (HCP) at \texttt{ www.humanconnectome.org}. Specifically, we used the database of healthy adult participants performing one of several motor tasks. Given a visual cue, the participants were asked to perform one of five tasks: tap left/right fingers, squeeze left/right toes, or move their tongue. Each block of the tasks follows a 3 sec cue and lasts for 12 sec. Each participant performed the experiment during two runs each consisting of 13 blocks. The acquisition used a repetition time of $0.72$ sec. The task is motivated by \cite{yeo2011organization} and \cite{buckner2011organization}. While the HCP implementation of the task is thoroughly described by \cite{barch2013function}, we provide a brief summary of data acquisition details. The whole-brain EPI images were acquired on a 3T Siemens Skyra machine with a flip angle at 52 degrees, BW = 2290 Hz/Px, collecting 72 slices with 2.0mm isotropic voxels, in-plane FOV = $208\times 180$mm.  In the motor-task, 284 frames per run were acquired. 

The SC tract counts were calculated using the Diffusion MRI Fiber Tracking tool in the DSI Studio software (``diffusion" as used in DTI, not the ``diffusion process" referred to elsewhere in this paper). Using a template image and a DTI diffusion scheme, we obtained the whole brain SC based on a total of 64 diffusion sampling directions. As a result 23,898 tracts were identified and their coordinates saved in the MNI space. From the resulting set of tracts the (pre-diffusion-process) SC was obtained from DSI Studio.

\subsection{Neuro-Hotnet}

To obtain the influence graph, we used the diffusion process with $\gamma = 30$, since this was the average node degree in the SC graph (this is the suggested approach for choosing $\gamma$ by \cite{vandin2011algorithms}). The diffusion process is robust to this choice since we found the choice of $\gamma = 1$ and $\gamma = 50$ both result in similar sets of estimated networks. The BOLD fMRI response was aggregated across N=308 participants by averaging each participant's sample correlation matrix. This alone constituted our observation-based connectivity, allowing for reducing the effect of outliers when incorporating a mixed set of participants and determining a consistent scale.

We choose a confidence level of $95\%$ ($\alpha = 0.05$) and $\delta = 0.155$ as this value of $\delta$ resulted in similar size networks as those from siGGM with Diffusion, allowing for a reasonable comparison between the areas of the brain identified by the different algorithms.

\subsection{siGGM with Diffusion}

A similar process was carried out using siGGM with Diffusion. Let $\tilde{\Sigma}$ denote the average of 308 participants' sample covariance matrices. The matrices $\tilde{\Sigma}$ and $G(\gamma)$ were used as inputs in the GGM framework, yielding the maximum a posteriori log-likelihood estimate of FC across all participants $C(\nu)$, with $\nu = 0.011$. All connected components $S(\nu)$ of $C(\nu)$ are shown in Table \ref{tab:resregions}. To recount, a small $\nu$ tends to result in fewer but larger components while a larger value results in numerous pairs and triplets, i.e. $\nu$ penalizes inclusion of edges, inducing sparsity and splitting discovered components into smaller groups. Small values of $\nu$ result in numerical issues in computation. $\nu = 0.011$ was chosen to find components of size around eight. This can be easily changed depending on practical and biological considerations.

\subsection{SC Naive methods}

% Power tests here. Note that my RunNaive.R script and RunHig.R with the "naive" parameter set to true yield very similar results, the latter using Higgin's parameter to turn on and off effect of SC.

Table \ref{tab:resregions} shows results for the SC naive approach with $\epsilon = 10^{-79}$. We selected this threshold to yield results of similar size to Neuro-Hotnet and siGGM for ease of comparison. The SC naive method partially agrees with the SC informed algorithms on the first component with very high tract and degree averages. The common nodes have high enough correlation in the fMRI data that all three algorithms discover them despite the down-weighting of influence during the diffusion process. This also serves to confirm that Neuro-Hotnet and siGGM can detect a highly correlated cluster in the fMRI data despite the down-weighting of influence. The second component in Table \ref{tab:resregions} has a tract average twice that (and a significantly higher degree average) of all other results from the SC informed methods.  The two most interesting discoveries in the context of motor network FC with low node degrees that the SC informed approaches agree on are not detectable with an SC naive approach.

% \begin{figure}[ht]
%     \centering
%     %used to be scale=0.6
%     \includegraphics[scale=0.6]{Figure9.pdf}
%     \caption{\nate{UPDATE THIS WITH NEW RESULTS} Euler diagram comparing discovered regions in Table \ref{tab:resregions} of size greater than 3 for all methods. Numbers in overlaps represent the number of regions in common between the components. The blue circles labelled ``H2,S2" - ``H5,S5" depict the perfect overlap of those regions for Neuro-Hotnet and siGGM with Diffusion.}
%     \label{fig:euler}
% \end{figure}

\subsection{Comparison of results}

In addition to the superior recovery rates found in Study 1, the SC-informed algorithms leveraging diffusion discover subnetworks with the HCP data that the SC naive method misses. The results of siGGM with Diffusion and Neuro-Hotnet agree closely (see Table \ref{tab:resregions}). S1 is a subset of H1 missing only one region, H3 and S3 are identical, and S4 and S5 together nearly comprise H5 with disagreement on two regions. One difference between the sparsity induced by $\nu$ and $\delta$ is that siGGM finds many more pairs. These results suggest that this use of siGGM may be better suited to detecting highly significant single edges and very small components. Neuro-Hotnet does better in only flagging components with low degree nodes, resulting in more interesting discoveries based on the philosophy described in subsection \ref{s:diff}.

\begin{figure}[ht]
    \centering
    \includegraphics[scale=0.6]{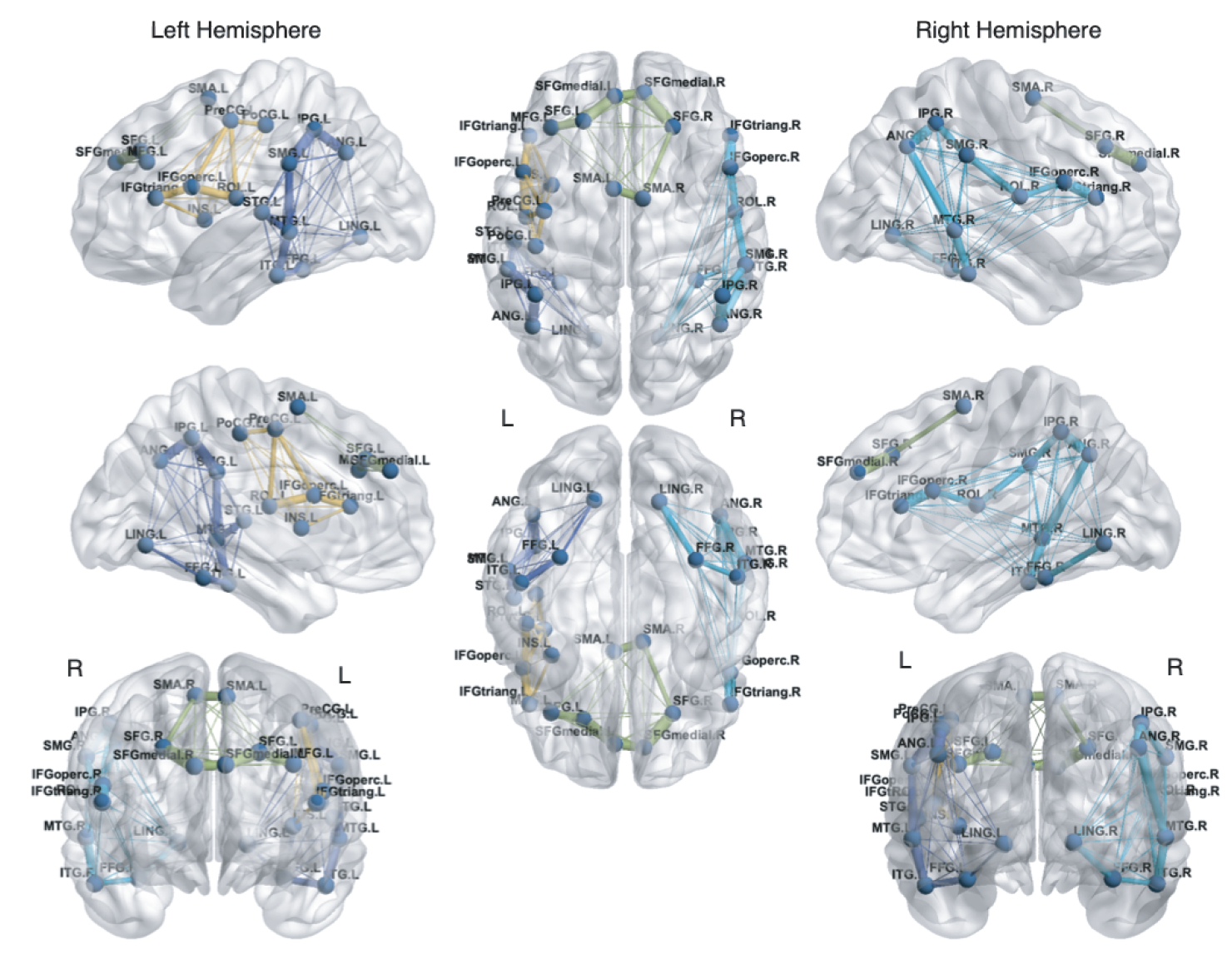}
    \caption{Lateral, Medial, Dorsal, and Ventral views of the components found with both Neuro-Hotnet and siGGM with Diffusion.}
    \label{fig:brains}
\end{figure}

Visualizations of the components found with both Neuro-Hotnet and siGGM are presented in Figure \ref{fig:brains}. Considering the functions of the networks obtained as a result of the proposed algorithms, the results reveal interesting associations between brain regions. For example, one of the subnetworks we obtain using the proposed Neuro-Hotnet approach involves the precentral gyrus, the site of the primary motor cortex, opercular and triangular parts of the inferior frontal gyrus associated with language processing; note that portions of the operculum include the secondary somatosensory cortex, and insula related to awareness, with all these regions occurring in the left hemisphere. The siGGM implementation yields  a similar network, though it also includes the postcentral gyrus or primary somatic sensory cortex. Interestingly, the SC naive approach does not include either the precentral or postcentral gyri, which would be expected as part of the findings given the performance of the motor task in this experiment.

\begin{table}
    \centering
   \begin{tabularx}{\textwidth}{|>{\centering\arraybackslash}p{0.1\textwidth}|p{0.04\textwidth}p{0.03\textwidth}Xp{0.1\textwidth}p{0.06\textwidth}|} \hline
    & \textit{Label} & \textit{Size} & \textit{Regions} & \textit{p-value} & \textit{Degree}\\ \hline
    
    \multirow{5}{*}{\shortstack{Neuro \\ Hotnet \\ $\delta \approx 0.155$}} & H1 & 5 & PreCG.L IFGoperc.L IFGtriang.L ROL.L INS.L & $<10^{-4}$ & 8,667\\ \cline{2-6}
    & H2 & 3 & PreCG.R PCL.L PCL.R & $<10^{-4}$ & 8,154\\ \cline{2-6}
    & H3 & 10 & IFGoperc.R IFGtriang.R ROL.R LING.R FFG.R IPG.R SMG.R ANG.R MTG.R ITG.R & $<10^{-4}$ & 5,647\\ \cline{2-6}
    & H4 & 5 & IFGorb .R OFCpost.R HES.R STG.R TPOsup.R & $<10^{-4}$ & 1,608\\
    \cline{2-6}
    % & \colorbox{neur}{H5} & 6 & OLF.L OLF.R REC.L REC.R OFCmed.L OFCant.L & $<10^{-4}$ & - & - \\
    % \cline{2-7}
    & H5 & 14 & PHG.L LING.L FFG.L IPG.L SMG.L ANG.L STG.L TPOsup.L MTG.L ITG.L Crus1.L Crus2.L CB4.L CB6.L & $<10^{-4}$ & 7,394\\
    \cline{2-6}
    & H6 & 3 & CB4.R Vermis4\textunderscore5.L Vermis6.R & $<10^{-4}$ & 1,722\\
    \hhline{|=|=====|}
    
    \multirow{8}{*}{\shortstack{siGGM \\ with \\ Diffusion \\ $\nu = 0.011$}} & S1 & 6 & PreCG.L IFGoperc.L IFGtriang.L ROL.L INS.L PoCG.L & - & 8,906\\ \cline{2-6}
    & S2 & 7 & SFG.L SFG.R MFG.L SMA.L SMA.R SFGmedial.L SFGmedial.R & - & 29,412\\ \cline{2-6}
    & S3 & 10 & IFGoperc.R IFGtriang.R ROL.R LING.R FFG.R IPG.R SMG.R ANG.R MTG.R ITG.R & - & 5,647\\ \cline{2-6}
    & S4 & 8 & LING.L FFG.L IPG.L SMG.L ANG.L STG.L MTG.L ITG.L & - & 6,657\\ \cline{2-6}
    & S5 & 6 & Crus1.L Crus2.L CB4.L CB6.L CB7.L CB8.L & - & 9,099\\ \cline{2-6}
    & S6 & 3 & Crus1.R Crus2.R CB6.R & - & 12,082\\ \cline{2-6}
    % 2 & 12 components & 6.43e-26 avg \\ \hline
    & - & 2 & 12 components & - & -\\ \hhline{|=|=====|}
    
    \multirow{4}{*}{\shortstack{SC Naive \\ $\epsilon = 10^{-79}$}} & N1 & 9 & SFG.L SFG.R MFG.L MFG.R IFGoperc.R IFGtriang.L IFGtriang.R SFGmedial.L SFGmedial.R & - & 24,195\\ \cline{2-6}
    & N2 & 10 & CAL.L CAL.R CUN.L CUN.R LING.L SOG.L SOG.R MOG. MOG.R IOG.L & - & 16,330\\ \cline{2-6}
    & N3 & 6 & SPG.L SPG.R IPG.L SMG.L PCUN.L PCUN.R & - & 5,899\\ \cline{2-6}
    & - & 2 & 6 components & - & -\\ \hline
    \end{tabularx}
    \caption{Discovered regions for both novel SC informed and the SC naive approach. Label annotates the regions as they are shown in Figure \ref{fig:brains}. Degree is the average number of tracts (in the raw DTI graph) ending in each node (weighted degree) over nodes in a given component. This reveals the final effect of the diffusion process in selecting for the interesting discoveries.}
    % \begin{tabularx}{\textwidth}{|p{0.1\textwidth}Xp{0.15\textwidth}|} \hline
    % \textit{Size} & \textit{Regions} & \textit{p-value} \\ \hline
    % 30 & $21,\ldots,50$ & 0.00e+00 \\ \hline
    % \end{tabularx}
    \label{tab:resregions}
\end{table}

\FloatBarrier
\section{Discussion}\label{s:discussion}

We found that Neuro-Hotnet exhibits comparable estimation accuracy to the commonly used current SC-informed FC estimation algorithms, while having superior performance compared to the existing methods in terms of computation time and scalability. The diffusion process appears promising and complements existing methods such as siGGM, leading them towards potentially more interesting discoveries. Neuro-Hotnet's biggest advantage relates to its simplicity of implementation and scalability, particularly when compared to other related methods. Through the combination of structural/functional data with few parameters and nonparametric hypothesis testing, every step of the process is easily reproducible and transparent. This makes Neuro-Hotnet fast to run, and easy to adapt to the rapidly changing field of FC estimation given its computational scalability. 

In addition to scalability, we showed that our proposed SC-informed approaches result in expected FC estimates given the task performed by study participants that were missed by commonly used existing approaches. Despite this high performance of Neuro-Hotnet, we believe that additional improvements can occur at every step of its implementation. First, it is worth considering other approaches for aggregating subject-level observations than averaging their cross-regional Pearson correlations. More sophisticated methods for combining data across participants may result in a more biologically meaningful aggregate. An approach to handling BOLD fMRI signals to take advantage of their temporal nature is the sliding-window approach, which could serve as an alternative to raw Pearson correlation computed over time for the full experimental data. Another alternative with less precedent is modeling the activation data over time as a multi-layer graph, where each time-point represents a distinct layer connected to adjacent time-point layers.  Pearson correlation can then be used to weight the edges on this graph. This representation allows the use of graphical methods to define observed influence between nodes. It is also possible to combine estimates across participants later in the algorithm. This may unfortunately come with significant run time costs depending on where in the algorithm this aggregation takes place. A strength of Neuro-Hotnet is that its structure allows for relatively easy modifications of the algorithm to incorporate these and other amendments and explore the effects of these changes on the results.

\section*{Data and code availability}

Data used in this project are publicly available for download from the HCP website. All appropriate Data Use Agreements were signed to obtain access to HCP data. We shared all the code used for simulations and data analyses on GitHub at \url{https://github.com/ntung88/NeuroHotnet}.

\section*{Acknowledgements}

This project was supported by National Institute of General Medical Sciences grant Number 5P20GM103645.

\section*{Conflicts of interest}

Declarations of interest: none.

\bibliography{neurohotnet}

\end{document}